\def\mehta{1}
\def\bgs{2}
\def\zycz{3}
\def\ott{4}
\def\stasprc{5}
\def\stasprep{6}
\def\stasprla{7}
\def\brody{8}
\def\towner{9}
\def\renyi{10}
\def\paladin{11}
\def\geisel{12}
\def\halsey{13}
\def\stasprlb{14}
\def\pc{15}
\def\plosz{16}

\font\titlefont=cmbx12

\magnification=\magstep1

\baselineskip=17pt

\vbox to 2.5cm {   }

\centerline{\titlefont SCALING PROPERTIES OF NUCLEAR GIANT}
\vskip\baselineskip
\centerline{\titlefont RESONANCE TRANSITION PROBABILITIES}

\vskip\baselineskip
\vskip\baselineskip
\centerline{Andrzej Z. G\'orski and S. Dro\.zd\.z}
\vskip\baselineskip
\centerline{Institute of Nuclear Physics, Radzikowskiego 152,
31--342 Krak\'ow, Poland}
\vskip\baselineskip
\centerline{\tt e-mail: \ gorski@alf.ifj.edu.pl}

\vskip\baselineskip
\vskip\baselineskip
\vskip\baselineskip

 Multifractal scaling analysis of nuclear giant resonance
transition probability distributions is performed within
the approximation which takes into account the one-particle--one-hole
($1p$--$1h$) and $2p$--$2h$ states.
A new measure to determine the fractal dimensions of the
spectra is introduced.
It is found that chaotic dynamics governing the decay leads to
non--trivial multifractal scaling properties. Such a kind of scaling
is absent in the case of regular dynamics.
The degree of collectivity is another element which
worsens the scaling.

\vskip\baselineskip


\vfill\eject

\centerline{\bf 1. Introduction}
\vskip\baselineskip

 There exist well established methods to identify chaotic behaviour
in classical non--linear dynamical systems.
It is, however, not fully clear what are the signatures of
chaoticity at the quantum level.
In fact, several indicators have been suggested.
For systems with complex energy spectra the statistical analysis of
their fluctuations has been introduced by Wigner in the context
of compound nuclei.
The Wigner form of the nearest neighbor spacing (NNS) distribution
turns out to be a good indication that
an underlying classical system is chaotic. This distribution
can be simulated within the random matrix theory
by the Gaussian orthogonal ensemble (GOE), while
the regular systems' spectra have Poissonian distribution
[\mehta, \bgs].
The NNS test can be successfully applied to
stationary systems and the information about the energy spectrum
is sufficient.
However, one would be interested to take into account more
information included in the wave function [\zycz].
The aim of this research is to find further signatures of chaos coming from the
structure of the wave functions.
In particular, chaotic dynamics on the classical level is known
to be associated with various fractal structures obeying the scaling
relations [\ott]. It would be extremely interesting to identify
a trace of such structures also on the quantum level.

To this end we shall consider the nuclear giant resonances
as physically interesting excitations and we shall investigate their
energy spectra and the transition probabilities
resulting from the coupling to more complex configurations.
These quantities are in principle measurable experimentally
thus our predictions are verifiable.
In particular, our calculations have been done for the $^{40}Ca$
nucleus and its $J^{\pi} = 2^+$ excitations to have the well defined
angular momentum and parity quantum numbers.
The centroid energies of nuclear giant resonances are relatively well
described
within the mean field 1-particle--1-hole ($1p$--$1h$) approximation.
To describe its decay one needs to include states of the $np$--$nh$
type. Truncating on the $2p$--$2h$ level is, however, technically
necessary (to have the hamiltonian matrix of a manageable size)
and physically well justified [\stasprc].
As the nuclear forces are predominantly two--body in nature
an initially excited giant resonance (a superposition of the $1p$--$1h$ states)
can couple directly to the $2p$--$2h$ states only [\stasprep].
In particular, within this approximation
one obtains a good agreement with experimental data.
Furthermore, the corresponding spectral fluctuations fulfill the GOE
characteristics [\stasprla] which implies an underlying chaotic dynamics and
the fact that already the subspace of the $2p$--$2h$ states properly
simulates relevant characteristics of the compound nuclei [\brody].

%%%\vfill\eject
\vskip\baselineskip
\vskip\baselineskip
\centerline{\bf 2. The physical model}
\vskip\baselineskip

 The general form of the Hamiltonian in our model reads:

$$
\hat H \ = \ \sum_i \epsilon_i \; a_i^\dagger a_i \ + \
{1\over2} \sum_{ij, kl} v_{ij,kl} \;
a_i^\dagger a_j^\dagger \; a_l \; a_k
\ , \eqno(1)
$$

\noindent where the first term ($H_0$) is the mean field part and the
second term ($V$) is the residual interaction.
For calculations we specify $H_0$ in terms of a local
Woods--Saxon potential including the Coulomb interaction.
The interaction part, $V$, was taken as the zero range Landau--Migdal
interaction with the empirical parameters taken from [\towner].

 The Hilbert space in our model is spanned by the
$1p$--$1h$ and $2p$--$2h$
vectors defined in terms of the creation and annihilation operators:

$$
\vert 1 \rangle \equiv a_p^\dagger a_h \vert 0 \rangle \ ,
\quad
\vert 2 \rangle \equiv a_{p_1}^\dagger a_{p_2}^\dagger
a_{h_2}  a_{h_1}  \vert 0 \rangle
\ , \eqno(2)
$$

\noindent such that they diagonalize the operator $H_0$ in
$1p$--$1h$ and $2p$--$2h$ sectors and
$\langle 1 \vert H_0 \vert 2 \rangle = 0$.

\noindent We prediagonalize the Hamiltonian $\hat H$ changing the
basis:

$$
\vert \tilde 1\rangle \equiv \sum_1 C^{\tilde 1}_1 \;
\vert 1 \rangle \ ,   \quad
\vert \tilde 2\rangle \equiv \sum_2 C^{\tilde 2}_2 \;
\vert 2 \rangle \ ,
\eqno(3)
$$

\noindent and the Schr\"odinger equation then reads:

$$
\left[ \matrix{ E_{\tilde 1} & A_{\tilde 1 \tilde 2} \cr
                 A_{\tilde 1 \tilde 2} & E_{\tilde 2} \cr}
\right]
\left[ \matrix{ X_{\tilde 1} \cr   X_{\tilde 2} \cr}
\right]
\ = \ E \
\left[ \matrix{ X_{\tilde 1} \cr   X_{\tilde 2} \cr}
\right]
\ , \eqno(4)
$$

\noindent where $E_{\tilde 1, \tilde 2}$ are diagonal block--matrices,
$A_{\tilde 1 \tilde 2}$, $A_{\tilde 2 \tilde 1}^\star$ are the
off--diagonal blocks and

$$
A_{\tilde 1 \tilde 2} = \sum_{12} C^{\tilde 1}_1 \;
\langle 1 \vert v \vert 2 \rangle \;
C^{\tilde 2}_2
\ . \eqno(5)
$$

\noindent The $A$ blocks are responsible for the coupling between these
two sectors and 'leaking' of the probability from the $1p$--$1h$
sector to the $2p$--$2h$ one.

 We consider two distinct and physically interesting cases:

\item{A).} No residual interaction in the subspace spanned by
$\vert 2 \rangle$ ($V=0$ in that subspace).
In this case the only non--vanishing matrix elements of the
type $\langle 2 \vert \hat H \vert 2^\prime \rangle$ are the diagonal
ones ($\langle 2 \vert \hat H \vert 2^\prime \rangle =
\langle 2 \vert \hat H_0 \vert 2^\prime \rangle= \epsilon_2
\delta_{22^\prime}$). The fluctuation properties of $\epsilon_2$
are then those characteristic for the regular systems [\stasprc].

\item{B).} All the matrix elements are included which results in the GOE
fluctuations of the energy spectra $\epsilon_2$ typical for classically
chaotic systems [\stasprc].

\noindent By comparing these two versions of our model one can then
study the influence of chaotic dynamics on quantum decay.

 The transition probabilities that will be used in the following
Section to define our measure (11) are taken as transitions between
the ground state ($\vert 0 \rangle$) and the excited states
($\vert i \rangle$) enumerated by $i=1, 2, \ldots, N$.
In our notation a state $\vert i \rangle$ can be expressed
as a superposition of states from both sectors:

$$
\vert i \rangle \ = \  \sum_1 a^i_1 \ \vert 1 \rangle +
\sum_2 a^i_2 \ \vert 2 \rangle
\ , \eqno(6)
$$

\noindent and the quadruple transition operator reads:

$$
\hat O \ = \ \hat Q \ Y_2(\Theta) \ r^2
\ , \eqno(7)
$$

\noindent where $\hat Q$ is the standard charge operator:
$\hat Q = {1\over 2} (1 + \tau_3)$, $\tau_3$ being the (isospin)
Pauli matrix. The first term in $\hat Q$ defines the isoscalar
and the second term
the isovector component of the transition.

 The transition probabilities can be computed directly
from the following formula:

$$
\rho_i = \vert \langle 0 \vert \hat O \vert i \rangle \vert^2
\ . \eqno(8)
$$

\noindent Since the transition operator defined by eq. (7) is one--body
in nature it picks-up the $1p$--$1h$ components of $\vert i \rangle$
only and thus:

$$
\rho_i \ = \ \big\vert \
\sum_{{\tilde 1} 1} X_{\tilde 1} \ C^{\tilde 1}_1 \
\langle 0 \vert \hat O \vert 1 \rangle
\ \big\vert^2
\  . \eqno(9)
$$

\noindent In principle, any state $\vert i \rangle$ includes certain admixture
of $\vert 1 \rangle$ and, therefore, an originally
(with no coupling to $\vert 2 \rangle$)
localized transition strength becomes much more fragmented.

 To have good quantum numbers we take the $^{40}Ca$ nucleus with the
angular momentum and parity: $J^\pi = 2^+$.
Taking into account four mean field nuclear shells
(two shells above and two below the Fermi surface)
this implies 26 $1p$--$1h$ states and 3~014 $2p$--$2h$ states.
Here, the centroid energy is 31 and 25 $MeV$ for isovector and isoscalar
transitions respectively, with dispersion of the order of 5 $MeV$.
Figs.~1 and 2 display the transition probabilities
for different energy levels in the regular (A) and chaotic case (B)
for isovector and isoscalar transitions, respectively.
Even at this level one can observe a clear distinction between
regular and chaotic cases. The later case even suggests a certain kind of
self-similarity regarding the clustering and the relative size of the
transitions. For this reason
the following Section is an attempt to perform a more systematic
analysis of these transition probabilities in the spirit of the Renyi
exponents [\renyi].

%%%%%%%%%%%%%%%%%%%%%%%%%%%%%%%%%%%%%%%%%%%%%%%%%%%%%%%%%%%%%%%%%%%%%%

\vskip\baselineskip
\vskip\baselineskip
\centerline{\bf 3. Scaling analysis}
\vskip\baselineskip

 The standard definition of a measure to compute the fractal
(Renyi) dimensions (exponents) for a set of $n$ points in $N$ boxes
(intervals) of a size $l$ is:

$$
p_i(l) \ = \  { n_i \over n }
\ , \eqno(10)
$$

\noindent where $n_i=n_i(l)$ denotes number of points in the $i$--th
box (interval) and the measure is properly normalized due to:
$\sum_i n_i = n$.
With this measure one can determine the fractal dimensions $d_q$
($q$ being any real number) by taking the limit $l \to 0$
and extracting the exponent from the following formula [\paladin]:

$$
\sum_i p_i^q(l) \ \sim \ l^{(q-1) d_q}
\qquad (l \to 0)
\ , \eqno(11)
$$

\noindent where the log--log plot of the left hand side
of eq. (11) {\it vs.} $N$ is used to compute the standard
scaling exponents $d_q$.

 Applying these ideas to our energy levels only
 we get not a very interesting result
as the energy spectrum of a typical quantum systems has
dimension equal to one and the same holds true in both our cases.
(There exist however systems whose energy spectra display
a fractal character [\geisel].)
Therefore, consistently with our previous discussion we are going
to consider
the sets of {\it pairs} of points, the $i$--th pair containing two numbers:
the energy and the transition probability, $\{ E_i, \rho_i\}$.
Taking into account the probabilities $\rho_i$ we have a more
complicated structure. Because to each energy corresponds exactly one
probability the whole structure is expected to have dimension in the range
$[0, \ 1]$ if the idea of scaling applies. Of course, due to non--uniformity
of the probability distributions the dimensions should depend on $q$ [\halsey].

 The specific value of $\rho_i$ can be interpreted as a frequency with
which the energy $\{E_i \}$ is 'visited' and this modifies an effective
number of energy levels in a given box.
In this way each energy point gets a
different weight. To have a measure with
the proper normalization we define, instead of (10),
the following new measure $P_i(l)$:

$$
P_i(l) \equiv
\left[ \ \sum_{\rm all\ } \ \rho_i \ \right]^{-1} \ \times
\sum_{E_i\in\ {\rm i-th\ box}} \; \rho_i
\eqno(12)
$$

\noindent where the summation in numerator goes over the probabilities
whose energies are included in the $i$-th box.
Here again the measure $P_i(l)$ is properly normalized:
$\sum_i P_i(l) = 1$.
The scaling exponent ("fractal dimension") with the measure
(12)
we will denote by $D_q$, while $d_q$ we reserve for the standard
fractal dimension. From (12)
it is clear that in the limit  $q\to0$ (capacity dimension)
the different nonzero probabilities $\rho_i$
give the same contribution to the left hand side of (11),
because the measure $P_i(l)$ is in the power of 0:

$$
\lim_{q\to0} P_i^q(l) \ = \ \lim_{q\to0} p_i^q(l) \ = \
\left\{ \matrix{
\ 1 \ \ \hbox{\rm if \ } i\hbox{\rm--th box is not empty,}\hfill\cr
\ 0 \ \ \hbox{\rm if \ } i\hbox{\rm--th box is empty.}\hfill    \cr
} \right.
\eqno(13)
$$

\noindent Hence, we have: $D_0 = d_0$, the last being the standard
capacity dimension.
In general, both measures and the corresponding dimensions differ for $q>0$
according to: $0 \le D_q \le d_q$. The differences reflect a degree of
non--uniformity in the probability distribution.
To determine the scaling exponents we use the formula (11) with the new
measure (12):

$$
\chi_q(l) \ \equiv \
\sum_i P_i^q(l) \ \sim \ l^{(q-1) D_q}  \qquad
(l \rightarrow 0) \ ,
\eqno(14)
$$

\noindent and the log--log plot of the quantity $\chi_q(l)$
of eq. (14) {\it vs.} the number of boxes $N$ is used to
extract the scaling exponents $D_q$ in Figs.~3 and 4.

 The input data (of Figs.~1 and 2) consist of the order of $\sim 2^{11}$
data points, the number sufficient
to display an exponential scaling but one should have in mind that
some statistical errors will be present, as the fractal dimension
formula involves the $l\to0$ (or, equivalently, the $N\to\infty$) limit.
In fact, for the chaotic case we have got a fairly good scaling
in the range of about 8 points in the log--log plot
as can be seen from Figs. 3(B). What is important,
this scaling persists up to the high
$q$ values, {\it i.e.} to the region where
the structure of the probabilities is mostly probed.
This kind of scaling has been considerably worsened in the regular case
(Fig. 3(A)~).
In the special case of the capacity dimension we get $D_0 \simeq 1$,
as in this case the scaling exponents are determined solely by the
energy distribution (see eq. (2)~). This is not a very interesting limit
and has not been plotted in Fig. 3(B).
 The regular case is also plotted for comparison, even though
the scaling soon degrades with increasing $q$  and choice of the scaling
exponents is difficult in this case. Hence, for $q>2$ the linear fits have not
been plotted in Fig. 3(A).
Also, as has been expected, we get larger differences between
$D_q$ and $d_q=1$ for greater $q$'s.

 Analogous plots for the isoscalar transitions are displayed in Fig.~4.
One can observe that in this case the scaling exponents (fractal
dimensions) are slightly (about 5\%) lower which is consistent with
the less uniform distribution of the corresponding transition
probabilities shown in Fig.~2.
Also, the scaling is less evident (its range is shorter, see Fig.~4(B)~)
even though the fluctuations in the subspace $2$ remain the same.
This may reflect the fact that the isoscalar giant resonance is more
collective\footnote{$^1$}{The notion of the collectivity used here
means localization of the transition strength in energy
and should not be confused for the same
notion used in the context of self--organization.}
than its isovector counterpart and as such it is
more resistive against decay [\stasprlb].
In other words the collectivity is expected to regularize the dynamics
and the result seems to confirm such a conjecture.

\vskip\baselineskip
\vskip\baselineskip
\centerline{\bf 4. Summary and conclusions}
\vskip\baselineskip

 In this paper we have investigated the scaling properties of
nuclear giant quadrupole resonance transition probabilities in
$^{40}Ca$ nucleus.
These quantities have been computed within the approximation
which includes the $1p$--$1h$ and $2p$--$2h$ states.
The results still preserve the Wigner form of the NNS distribution of energy
levels and they are in good agreement with the experimental data
(see Figs.~1, 2 and [\stasprc, \stasprla]).

 To estimate the scaling properties we have introduced
a measure defined by (12) which combines the energy spectra and
the transition
probabilities. This definition allows to make use of the concept of the
generalized Renyi exponents [\renyi, \paladin].
Based on this concept it has been shown that for both,
the isovector and the isoscalar transitions,
one can speak about the scaling exponents $D_q$ of the multifractal type.
This observation applies, however, only to the case when the physics of
fragmentation is governed by the chaotic dynamics.
These results can be treated as an interesting indication of what are
the further signatures of classical chaos on the quantum level.

For the isovector transitions the scaling is somewhat better and
the exponents are about 5\% higher.
This effect may have to do with the fact that the isoscalar resonance
is more collective than its isovector counterpart.
The collectivity is a natural element regularizing the dynamics.

 It has been suggested [\pc] that dynamical systems of the type discussed
in this paper can be simulated by the binary, self--similar and
conservative
random fragmentation process which yields universal behaviour
independent of the precise fragmentation mechanism [\pc, \plosz].
This gives another justification for neglecting the higher order
excitations.

\vskip\baselineskip
 We thank Marek P\l oszajczak for very
useful discussions.
This research was supported by KBN Grant 2 P03B 140 10.

\vfill\eject
\vskip\baselineskip
\vskip\baselineskip

REFERENCES
\vskip\baselineskip

\item{[\mehta]} M.L. Mehta, {\it Random Matrices and the Statistical
Theory of Energy Levels} (Academic, New York, 1967).

\item{[\bgs]} O. Bohigas, M.J. Giannoni and C. Schmit, {\it Phys. Rev.
Lett.} {\bf 52}, 1 (1984).

\item{[\zycz]} K.~\.Zyczkowski, {\it Acta Phys. Pol.} {\bf B24}, 967
(1993).

\item{[\ott]} E.~Ott, {\it Chaos in dynamical systems}
(Cambridge University Press 1993).

\item{[\stasprc]} S. Dro\.zd\.z, S. Nishizaki, J. Speth and J. Wambach,
Phys. Rev. {\bf C49}, 867 (1994).

\item{[\stasprep]} S.~Dro\.zd\.z, S.~Nishizaki, J.~Speth and J.~Wambach,
{\it Phys. Rep.} {\bf 197}, 1 (1990).

\item{[\stasprla]} S. Dro\.zd\.z, S. Nishizaki and J. Wambach,
{\it Phys. Rev. Lett.} {\bf 72}, 2839 (1994).

\item{[\brody]} T.A. Brody, J. Flores, J.B. French, P.A. Mello, A. Pandey
and S.S.M. Wong, {\it Rev. Mod. Phys.} {\bf 53}, 385  (1981).

\item{[\towner]} J.~S.~Towner, {\it Phys. Rep.} {\bf 155}, 263 (1987).

\item{[\renyi]} J.~Balatoni and A.~Renyi, {\it Publ. Math. Inst. Hung.
Acad. Sci.} {\bf 1}, 9 (1956)
[also in: {\it Selected papers of A. Renyi}, Vol. 1
(Academia Budapest 1976), p. 558].

\item{[\paladin]} G.~Paladin and V.~Vulpiani,
{\it Phys. Rep.} {\bf 156}, 148 (1987).

\item{[\geisel]} T. Geisel, R. Ketzmerick and G. Petschel,
{\it Phys. Rev. Lett.} {\bf 66}, 1651 (1991).

\item{[\halsey]} T.C. Halsey, M.H. Jensen, L.P. Kadanoff, I Procaccia
and B.I. Shraiman, {\it Phys. Rev.} {\bf A33}, 1141 (1986).

\item{[\stasprlb]} S.~Dro\.zd\.z, S.~Nishizaki, J.~Wambach and J.~Speth,
{\it Phys. Rev. Lett.} {\bf 74}, 1075 (1995).

\item{[\pc]} A. Z. G\'orski, R. Botet, S. Dro\.zd\.z  and
M. P{\l}oszajczak,
{\it Proceedings of the 8th Joint EPS--APS International Conference
on Physics Computing}   (September  1996), to appear.

\item{[\plosz]} R.~Botet, and M.~P\l oszajczak, {\it Phys. Rev. Lett.}
{\bf 69}, 3696 (1992);
{\it J. Mod. Phys.} {\bf E3}, 1033 (1994).

\vfill\eject
\vskip\baselineskip
\vskip\baselineskip
\centerline{\bf Figure captions.}
\vskip\baselineskip

\item{Fig. 1} Isovector quadrupole strength distribution in $^{40}$Ca
for the regular (A) and the chaotic (B) case. Note different energy scales.
\item{Fig. 2} The same as Fig.~1 for the isoscalar strength distribution.
\item{Fig. 3} The log--log plot of $\chi_q(l)$ of eq. (14)
{\it vs.} the number
of boxes $N$ (equivalent to the inverse of the box size $l$).
$\chi_g(l)$ is determined by the isovector quadruple transition
probabilities in the regular (A) and chaotic (B) case for
$q=2,4,6$.
\item{Fig. 4} The same as Fig.~3 for the isoscalar transition
probabilities.

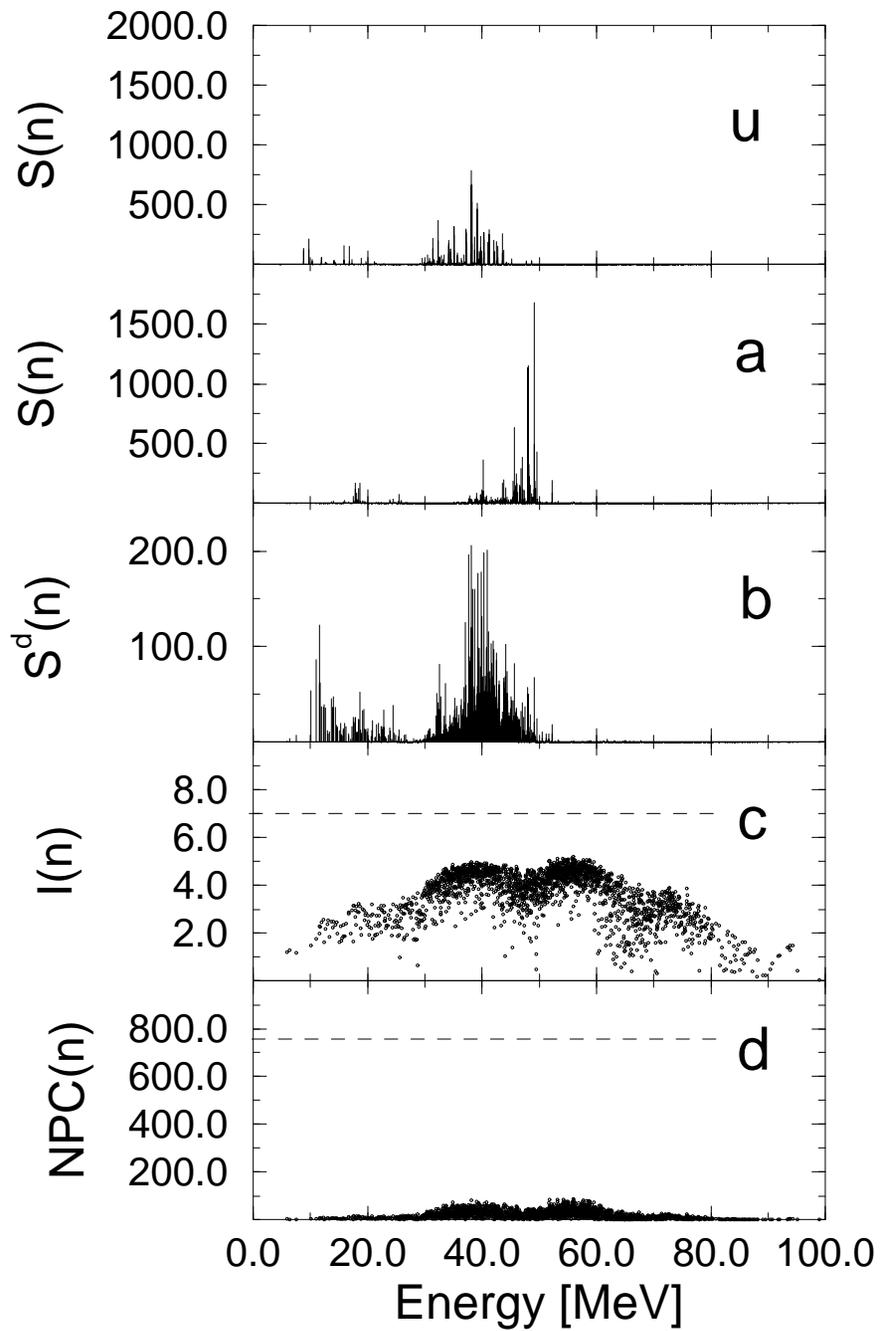
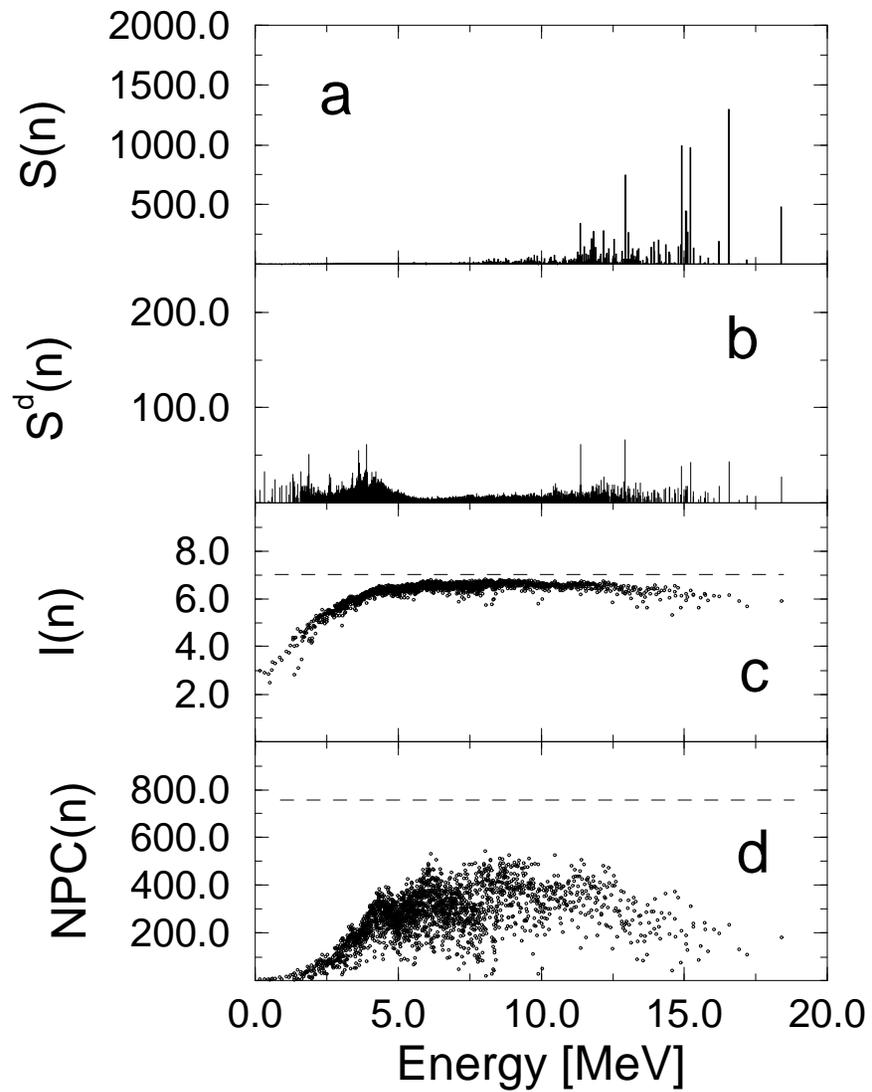
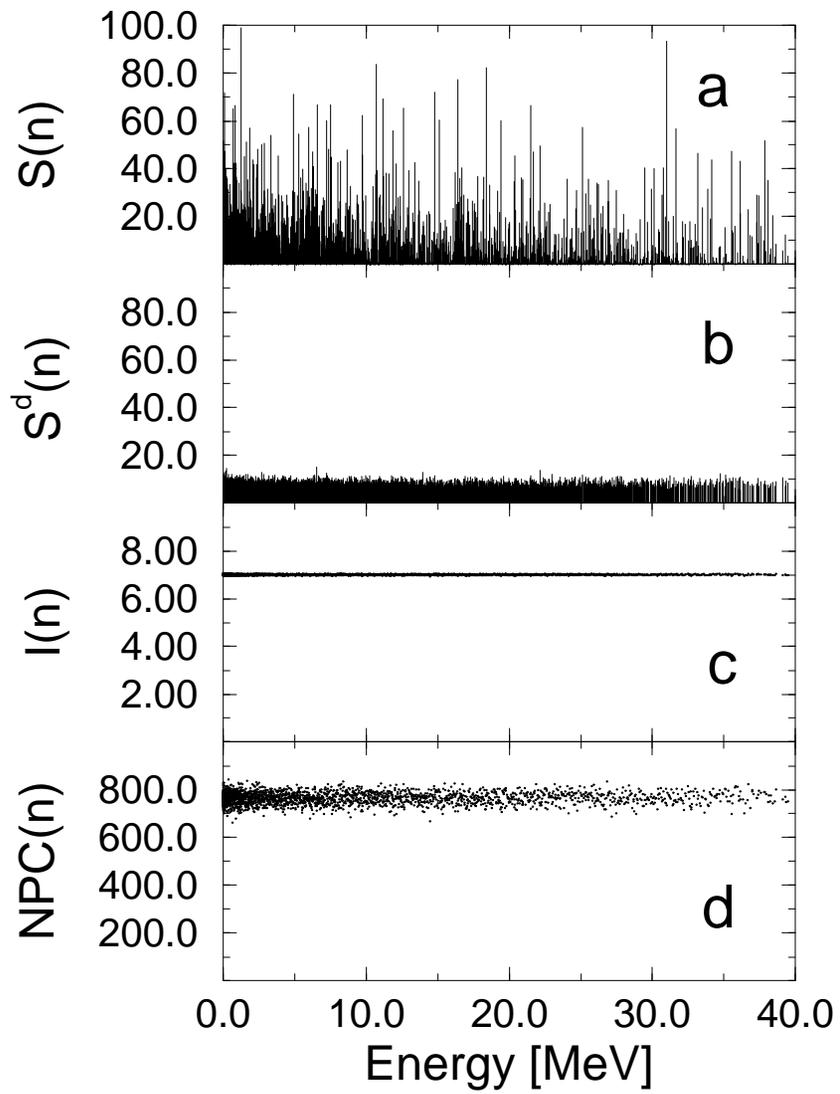
\bye